\documentclass{svmult}
\usepackage[latin1]{inputenc}
\usepackage[english]{babel}
\usepackage{graphicx}
\usepackage{cite}
\usepackage{makeidx}
\usepackage{multicol}
\usepackage{verbatim}
\usepackage{amssymb}
\usepackage{subfigure}
\usepackage[sumlimits]{amsmath}
\usepackage{xspace}

\DeclareGraphicsExtensions{.eps,.epsi,.ps, .png}

\makeindex

\title*{Changing levels of description in a fluid flow simulation}
\author{
	Pierrick Tranouez\inst{1} 
	\and 
	Cyrille Bertelle\inst{1} 
	\and 
	Damien Olivier\inst{1}}
\institute{
	LITIS
	25 rue Philippe Lebon
	BP 540 76058 Le Havre Cedex
	France
	\texttt{[FirstName.Surname]@univ-lehavre.fr}}

\begin{document}
\graphicspath{{./eps/}{./}{./Crobars/}{../Figures/}{../images/}}
\maketitle

\begin{abstract} 
\index{Complex system}
We described here our perception of complex systems, of how we feel the different layers of description are an important part of a correct complex system simulation. We described a rough models categorization between rules based and law based, of how these categories handled the levels of descriptions or scales. We then described our fluid flow simulation, which combines different fineness of grain in a mixed approach of these categories. This simulation is built keeping in mind an ulterior use inside a more general aquatic ecosystem.

\end{abstract}

\section{Introduction}
\label{sec:Introduction}
\index{Complex system}
\index{Simulation}
\index{Complex system!Living system}\index{Complex system!Living system!Ecosystem}.
Not everyone agrees on what a complex system is, but many authors agree on some properties of these objects, among them the difficulty of their simulation. We will describe here some of the causes we perceive of this difficulty, notably the intricate levels of description necessary to tackle their computer simulation \index{Scale}. We will then explain how this notion guided our simulation, a fluid flow computation with solid obstacles, meant as a step toward a broader simulation of an aquatic ecosystem 

\section{Simulation of complex systems}
\label{sec:Simulation}

\subsection{Complexity vs reductionism}
\label{sec:Complexity}
As explained in \cite{Bertalanffy1968}, to study how mass works in a material system, dividing this system into smaller parts is a good method. Indeed, each of his subsystems is massive, and therefore the study, the reductionism, \index{Simulation!Reductionism} can continue.

This is not so with living systems. If you divide a living system into smaller parts, the odds are good that all you reap is a heap of dead things. That's because the life question of a live system is \emph{complex}. This means that what is important is not so much the parts of the systems, nor the parts of these parts, but the functioning relations that exist between them.

This is one of the two main reasons why one may want to integrate the multiple possible scales of description into a simulation. When you enquire about a complex question in a system, you need to choose carefully the needed levels of description, as you can't simplify them. Furthermore, these needed levels may change during the simulation, and it would be a fine thing if the simulation could adapt to these variations.\\
Thus changing the scales of description during the simulation could be useful for the accuracy of the answers to complex questions regarding the system the simulation may provide \index{Simulation!Multi-scale|see{Scale}}.

Then, there is the understanding of these answers.

\subsection{Clarity of the simulation}
\label{sec:Clarity}
Users of a simulation \emph{question} it \index{Simulation}. Final users ponder about the future of the thing simulated in various circumstances, developers try to ascertain the validity of their model and of its implementation, but all use it with a purpose in mind.

Choosing the right level of description \index{Level of description|see{Scale}}is then important to give a useful answer. If the simulation is able to adapt its descriptions to what is needed by its user, lowering the noise and strengthening the signal, by choosing the right level(s) of description, it will be a better tool. For example in our application, a simulation of a fluid flow in an ecosystem, this help takes the form of hiding tiny perturbation and putting forward the main structures of the flow that emerged during the simulation.

\section{Methods for changing the scale in a simulation}
\label{sec:Methods}

\subsection{Law-based vs. rule-based models}
\label{sec:Law}
Classifying the various ways science can tackle problems is an arduous task. We will nonetheless distinguish two rough categories of models.\\
\index{Simulation!Law-based model}
Law-based models are the most used in science, most notably in physics. They are often continuous, especially in their handling of time and space, and based on a differential formulation whose resolution, ideally formal but often numerical, computes the values of state variables that describe the studied domain. Those methods are sometimes also called \emph{global} or \emph{analytical}.

In rule-based models \index{Simulation!Rule-based model}, the studied domain is discretized in a number of entities whose variations are computed with rules. There is therefore no longer a global description of the domain, nor is there a priori continuity. Cellular automata fall in this category of course \index{Simulation!Cellular automata}, and so do objects/actors/agents \index{Simulation!Objets} \index{Simulation!Actors} \index{Simulation!Agents}. Those models have had a strong influence on game theory, and from there directly on social models, and later on other domain through computer science for instance, at least by way of metaphors. Those models have other names depending on the domain where they are used, ranging from \emph{micro-analytical} in sociology, to \emph{individual-based} \index{Simulation!Individual Based Model} in life sciences or just simply \emph{local}.

Both kinds of models can be deterministic or stochastic. Finally, to blur the distinctions a bit more, models may include sub-parts falling in any of these categories. This is often the case with ecosystems for instance.

\subsection{Changing the scale in law-based models}
\label{sec:ChangingLaw}
\index{Scale} \index{Simulation!Law-based model} 
Accessing different levels of description in these models is often done through integration. Indeed, as said before, state function in these models are often continuous, and can therefore be integrated. New state functions are then valued or even built, on another domain and based on different phenomenological equations. For example, A. Bourgeat
\cite{Bourgeat1997} describes fluid flows in porous milieus, where, from Navier-Stokes equations, through integration and the addition of an extra parameter, he builds a Darcy law. These changes of equations description from one level to another alter sometimes drastically the linearity of the models and may lead to the introduction of new parameters that act as a memory of the local domain inside the global one.

In a similar way to this example, the change of level of description in analytical models is often performed \emph{a priori}, at the building of the model.

\subsection{Changing the scale in rule-based models}
\label{sec:ChangingRule}
\index{Scale}\index{Simulation!Rule-based model} 
Models based on rules offer a wider variety of ways of changing the levels of description. Indeed, local approaches are better designed to integrate particularities of very different entities and their mutual influence, as is the case when entities of various scales interact.

\subsubsection{Cellular automata}
\label{sec:CellularAutomata}
\index{Simulation!Cellular automata}
The first individual based computer science structures may have been cellular automata. If they were created by Stanislas Ulam, Von Neumann self-replicating automata may have been the foundation of their success \cite{VonNeuman1966}. Ulam himself already noticed that complex geometric shape could appear starting with only simple basic blocks. Von Neumann then Langton \cite{Langton1986} expanded this work with self-replicating automata.

If shapes and structures did appear in the course of these programs, it must be emphasized that it were users, and not the programs themselves, that perceive them. Crutchfield \cite{Crutchfield1992} aimed at correcting that trend, by automating the detection of emergent structures.

Detecting structures has therefore been tried, but reifying these structures, meaning automatically creating entities in the program that represent the detected structures has not been tackled yet, as far as cellular automata are concerned. It could be that the constraint on its geometry and the inherent isotropy of the cellular automata are in this case a weakness.

\subsubsection{Ecology}
\label{sec:Ecology}
\index{Ecology}
Since the beginning of the use of individual based models in ecology, the problem of handling the interactions between individuals and populations occurred \cite{DeAngelis1992}. The information transfers between individual was handled either statistically \cite{Caswell1992} or through the computing of action potential \cite{Palmer1992}.

\subsubsection{DAI uses}
\label{sec:DAI}
Most software architectures designed to handle multiple levels of description are themselves hierarchical. They often have two levels, one fine grained and the other coarse grained. Communication between these two levels could be called decomposition and recomposition, as in \cite{Marcenac1997}.

\index{Complex system!Living system!Ecosystem}
\index{Fluid flow}
In 1998, members of the RIVAGE project remarked that it was necessary in multi-agent simulations, to handle the emergent organizations, by associating them with behaviors computed by the simulation \cite{Servat1998}. Before that, were handled only border interactions between entities and groups \cite{Gasser1992}.

This led in D. Servat PhD thesis to a hydrodynamic model incorporating in part these notions. In his Rivage application, water bowls individuals are able to aggregate in pools and rivulets.  The individuals still exist in the bigger entities. The pros are that it enables their easily leaving the groups, the cons that it doesn't lighten in any way the burden of computing. Furthermore, these groups do not have any impact on the trajectories of the water bowls.

\section{Application to a fluid flow}
\label{sec:FluidFlow}
\index{Fluid flow}
\subsection{Ontological summary}
\label{sec:Ontological}
The fluid flows that constitute the ocean currents on the planet are the result of an important number of vortexes of different scales. Turbulent movement can also be decomposed into vortexes, on scales going down to the near molecular. Viscosity then dissipates kinetic energy thus stopping the downward fractal aspect of these vortexes \cite{Lesieur1987}. There are qualitatively important transfers of energy between these various scales of so different characteristic length.  Representing these is a problem in classic modeling approaches.

In classic, law based models, turbulent flows are described as a sum of a deterministic mean flow and of a fluctuating, probabilistic flow. These equations (Navier-Stokes) are not linear, and space-time correlation terms must be introduced to compensate for that. These terms prevent any follow up of the turbulent terms, and thus of the energy they transmit from one level to another.

A pure law based approach is therefore not capable of a qualitative analysis of the transfer of energy between the different scales of a turbulent flow. A multi-level model, where multiple scales of vortexes would exist, and where they would be able to interact, would be a step in this qualitative direction.

\subsection{Treatment of multiple scales}
\label{sec:MultipleScales}
\subsubsection{Fluid mechanic model and its structures}
\label{sec:FluidMechanic}
\index{Fluid flow}
\index{Fluid flow!Vortex method}
There are a number of models used to describe fluid flows. The set we use here are based on a discretisation of the flow, and are called vortex methods \cite{Leonard1980}.

In vortex methods, the flow is separated in a number of abstract particles, each being a local descriptor of the flow. These particles indicate the speed, vorticity etc \ldots of the flow where they are located.\\
These particles are not fixed: they are conveyed by the fluid they describe.\\
We find this model interesting as it is a local model, hence better able to deal with local heterogeneities. The values of the properties the particles describe are computed through the interactions between the particles, most notably through Biot-Savart formula. More details on this computation can be found in \cite{Bertelle2000}.

The vortex method we use is of $O(n^2)$ complexity. Finding ways of lightening this calculus is therefore important. One lead is through making our model multi-scale, and only computing entities at the scale we need them. This is our second motivation for our using different levels of description.
In order to have different levels of description, we will have to use an adapted description of the simulation entities.\\
These entities come and go during the simulation, and thus we need a method to change the level of their description \emph{during} the simulation, and not beforehand the way it is usually done.

In our fluid flow, the main entities as we explained are vortexes. Not only do we therefore need to detect emerging vortexes by monitoring lower level vortexes particles, but also, as these vortexes aggregate among themselves to form even bigger vortexes, make this detection process iterative.
Detecting the structures is not enough: we also need to create them in the simulation once they are detected. We must make these new entities live in the simulation, interacting with its various inhabitants (most notably particles, vortexes). They must evolve, whether it is growing or decaying to its possible disintegration.

Let us now describe our recursive detection-creation/evolution-destruction cycle.

\subsubsection{Detecting emergent vortexes among the vortex particles}
\label{sec:Detection}
\index{Emergence}
Structures are detected as clusters of particles sharing some properties. For vortexes these properties are spatial coordinates and rotation sense.
As described in the following figure (figure \ref{fig:DSE}), the process is:
\index{Delaunay triangulation}
\begin{enumerate}
	\item Delaunay triangulation of the particles
	\item Computation of a minimal spanning tree of this triangulation
	\item Edges that are too much longer than the average length of edges leading to the particles are removed. 
	So are edges linking particles of opposite rotational.
	\item The convex hull of the remaining trees is computed
	\item An ellipse approximates the hull through a least square method. Geometric constraints on the proportions 
	of the ellipse must be satisfied or the vortex is not created.
\end{enumerate}
Further details on this process can be found in \cite{Tranouez2003}.

\begin{figure}
	\centering
	\includegraphics[width=\textwidth]{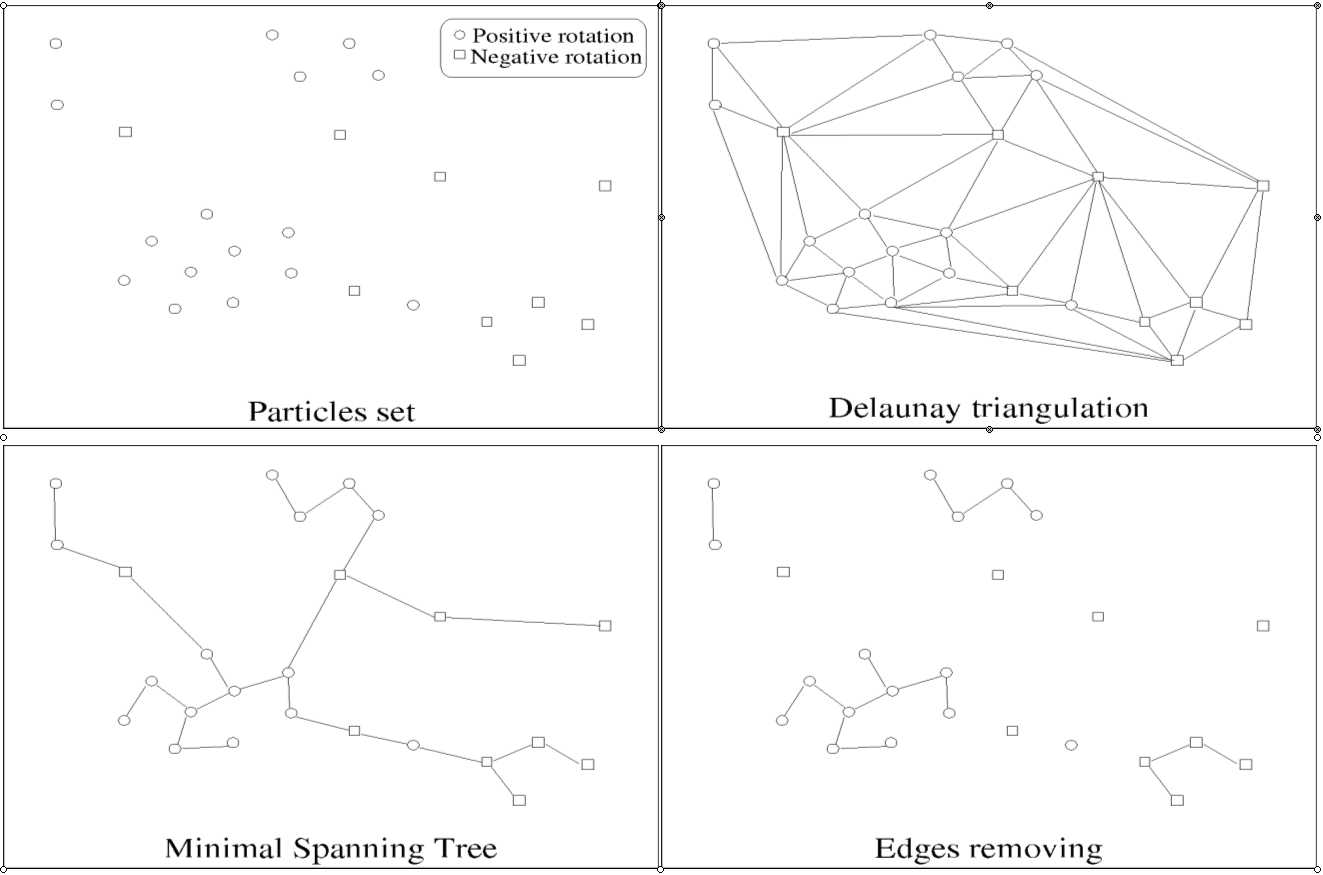}
	\caption{Basic detection scheme}
	\label{fig:DSE}
\end{figure}

\subsubsection{Scale transfer : making simulation entities of the detected structures}
\label{sec:ScaleTransfer}
\index{Automata}
\index{Multiplicity automata|\see{ Automata }}
\index{Transducer|\see{ Automata }}
\index{Eco-resolution}
Detected structures are created in the simulation where they take the place of the particles whose interactions gave them birth.\\
The vortex structures are implemented through multiplicity automata \cite{Bertelle2001}. These automata handle both the relations between higher level vortexes and the relations between them and the basic particles.

Part of the relations between vortexes and their environment are handled through a method based on the eco-resolution model \cite{Drogoul1992}, in which entities are described through a perception and combat metaphor.

\begin{figure}
	\centering
	\includegraphics[width=\textwidth,bb=0 0 640 476]{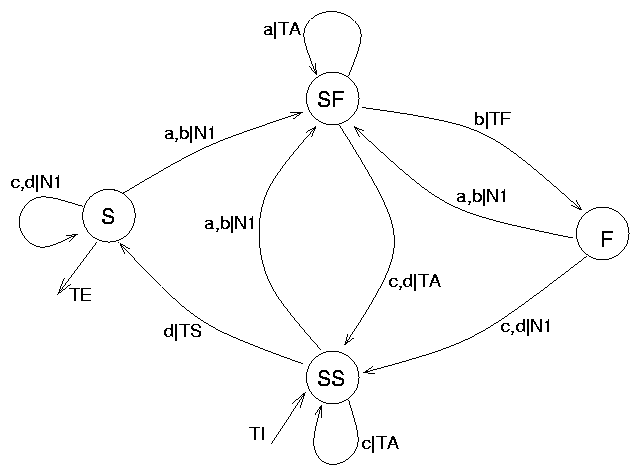}
	\caption{Transducer managing the stability of a vortex}
	\label{fig:Transducer}       
\end{figure}

The associated perceptions and actions are:
\begin{enumerate}
	\item	Perceiving an intruder means being on a collision course with another vortex. Figure 3 sums up the 
	various possibilities of interception by vortexes of opposed rotation and how each is translated in a 
	perception.

	\begin{figure}
		\centering
		\includegraphics[width=\textwidth,bb=0 0 1003 767]{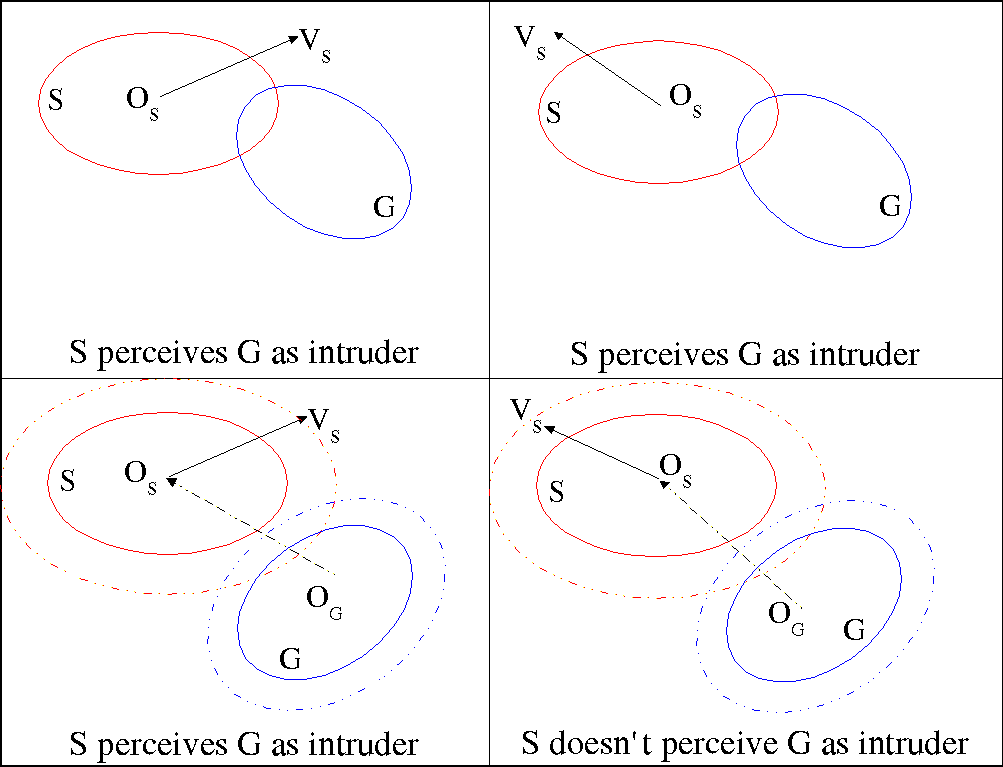}
		\caption{Perception of an intruder}
		\label{fig:Intruder}
	\end{figure}

	\item	Attacking another vortex means sending it a message.
	\item	Being attacked means receiving such a message.
	\item	Fleeing means being destabilized: the vortex structure shrinks and creates particles on its border. 
	(Figure 4). Too much flight can lead to the death of the structure, which is then decomposed in its basic 
	particles.

	\begin{figure}
		\centering
		\includegraphics[width=\linewidth,bb=0 0 329 102]{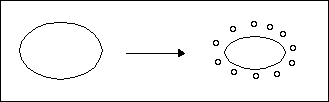}
		\caption{Destabilization}
		\label{fig:Destabilization}       
	\end{figure}

	\item	Getting satisfaction can mean two things. One is aggregating surrounding particles of compatible 
	vorticity. This calculation is done through a method close to the initial structure detection: Delaunay 
	triangulation, spanning tree, removal of edges.  Compacity criteria are then used to estimate whether the tree 
	should be added to the vortex and thus a new ellipse be computed or not. For instance in Figure 5, the 
	particles on the lower left will be aggregated while those on top will not. The other is fusing with a nearby 
	vortex of  the same rotation sense, if the resultant vortex satisfies compacity geometric constraints (it 
	mustn't be too stretched, as real fluid flow vortexes are not).

	\begin{figure}
		\centering
		\includegraphics[width=\textwidth, bb= 0 0 745 510]{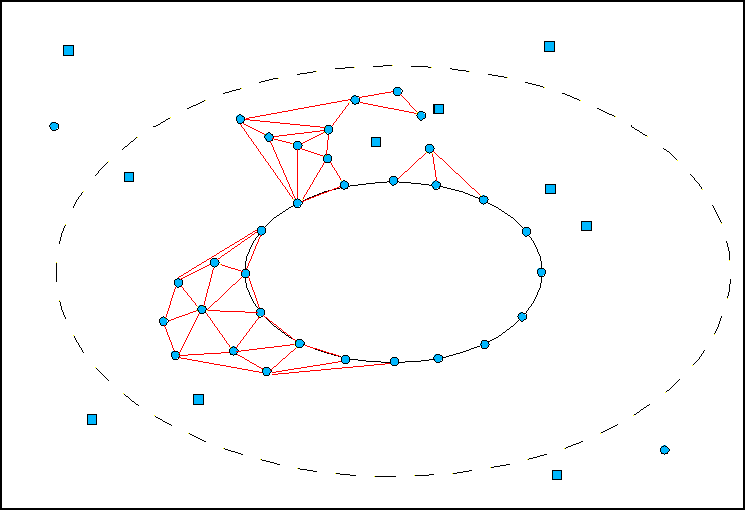}
		\caption{To aggregate or not to aggregate}
		\label{fig:AggregateOrNot}       
	\end{figure}

\end{enumerate}

The Eco-agent manages the stability of the structures. Its trajectory behavior depends on adapted Biot-Savart formula.

The described process is then iterated. New structures are detected and implemented, while others grow, shrink or disappear altogether. They move according to fluid mechanics laws. What remains to be seen is: how do they interact with solids?

Further details on this process can be found in \cite{Tranouez2003}.

\subsection{Interaction with solids}
\label{sec:Interaction}
\index{Fluid flow!Interactions}
Solids in a fluid flow simulation are a necessary evil, acting as obstacles and borders. Their presence requires a special treatment, but their absence would prevent the simulation of most interesting cases. They bring strong constraints in their interactions with the fluid, and as such introduce the energy that gives birth to the structures we're interested in.

\begin{figure}
	\centering
	\includegraphics[width=\textwidth]{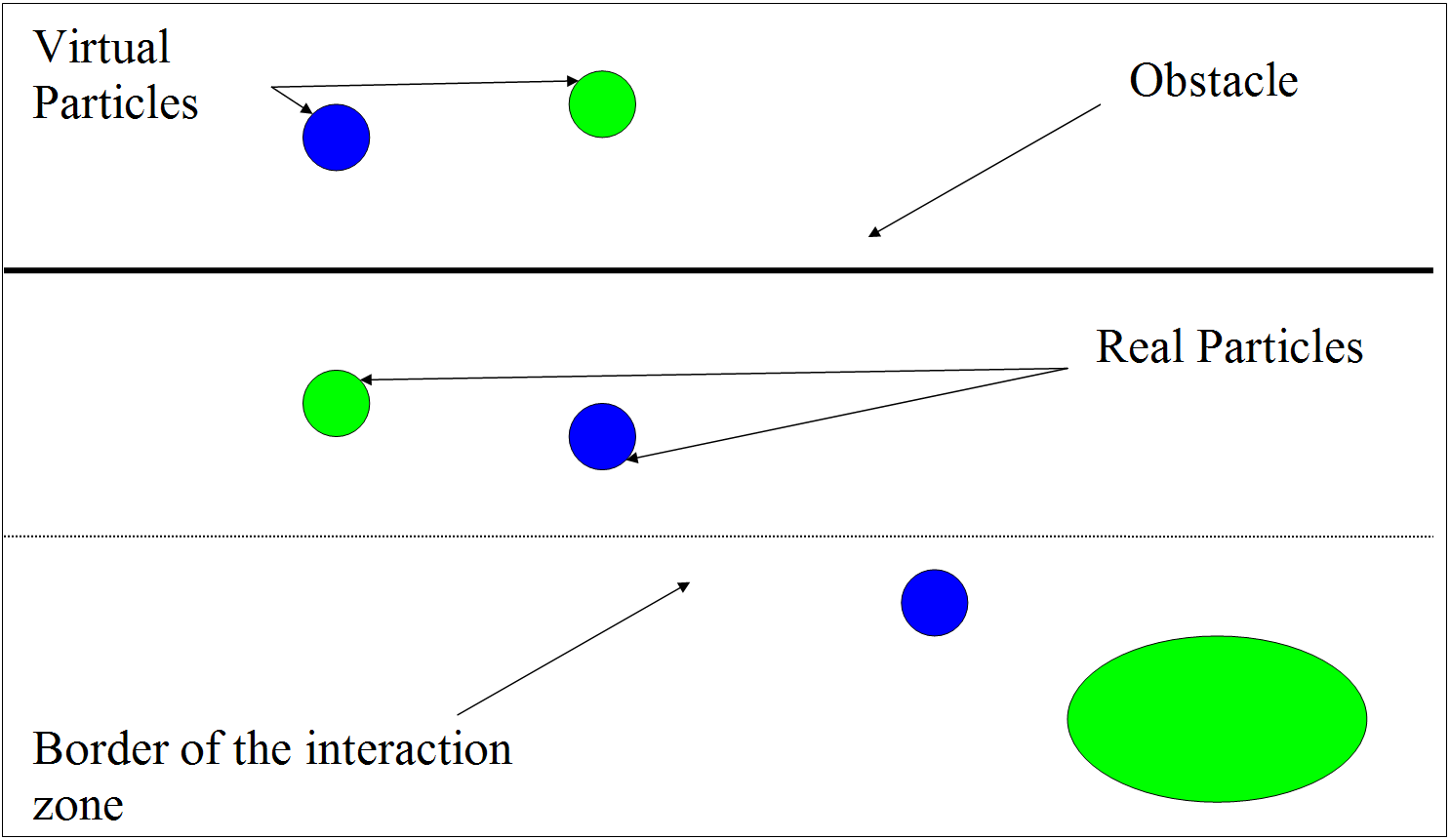}
	\caption{Plane obstacle (e.g. border)}
	\label{fig:Obstacle}
\end{figure}

\begin{figure}
	\centering
	\includegraphics[width=\textwidth]{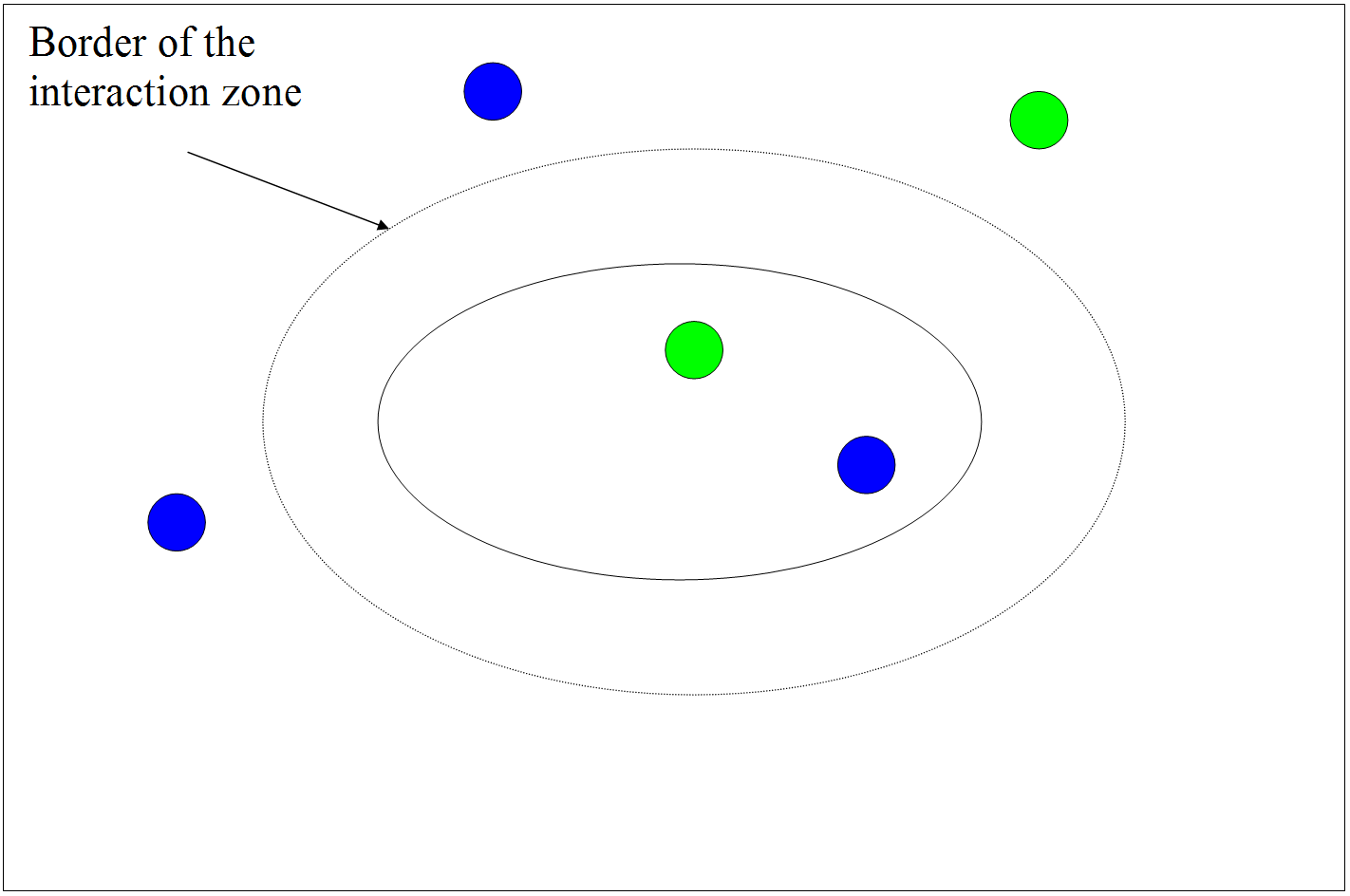}
	\caption{Elliptic obstacle}
\label{fig:Elliptic}       
\end{figure}

We manage structures using virtual particles. A solid close enough to real particles generates these particles symmetrically to its border (cf. Figure \ref{fig:Obstacle} and Figure \ref{fig:Elliptic}). The real particles perceive the virtual particles as particles in the Biot-Savart formula. The virtual particles on the other hand do not move. They are generated at each step if necessary, at the right place to repel the particles the obstacle perceived.

Obstacles also generate virtual particles to try to repel structures. If it fails, obstacles aggress the vortex, thus making it shrink so as to deal with the interaction solid/fluid in a finer grain.

\begin{figure}
	\centering
	\includegraphics[width=\textwidth]{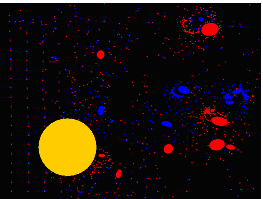}
	\caption{Simulation screenshot Red and blue : particles and vortexes Yellow disc : Obstacle In case of shades 
	of grey, the disc is at the bottom left, surrounded by the fluid}
	\label{fig:Screenshot}       
\end{figure}

\section{Conclusion}

We described here our perception of complex systems, of how we feel the different layers of description are an important part of a correct complex system simulation. We described a rough models categorization between rules based and law based, of how these categories handled the levels of descriptions or scales. We then described our fluid flow simulation, which combines different fineness of grain in a mixed approach of these categories. This simulation is built keeping in mind an ulterior use inside a more general aquatic ecosystem.

Our result show so far an adequate real time handling of the interactions between fluids and solids. Other methods can statically simulate in better details this interaction, but it often requires a knowledge beforehand of the placement of the solids. We can deal with solids moving at random, or more interestingly computed by the simulation itself, and not known before the simulation begins. We hope it will enable us to simulate for example animals or dynamic obstacles, thus integrating our more global works on ecosystem simulation \cite{Tranouez2005}.


\bibliographystyle{plain}
\bibliography{epnads}

\begin{thebibliography}{10}

\bibitem{Bertalanffy1968}
L.~Von Bertalanffy.
\newblock {\em General System Theory: Foundations, Development, Applications}.
\newblock George Braziller Inc, New York, 1968.

\bibitem{Bertelle2001}
C.~Bertelle, M.~Flouret, V.~Jay, D.~Olivier, and J.L-Ponty.
\newblock Automata with multiplicities as behaviour model in multi-agent
  simulations.
\newblock In {\em SCI'2001}, 2001.

\bibitem{Bertelle2000}
C.~Bertelle, D.~Olivier, V.~Jay, P.~Tranouez, and A.~Cardon.
\newblock A multi-agent system integrating vortex methods for fluid flow
  computation.
\newblock In {\em 16th IMACS Congress}, volume 122-3, Lausanne (Switzerland),
  August 21-25 2000.
\newblock electronic edition.

\bibitem{Bourgeat1997}
A.~Bourgeat.
\newblock {\em Tendances nouvelles en mod\'elisation pour l'environnement},
  chapter Quelques probl\`emes de changement d'\'echelle pour la mod\`elisation
  des �\'ecoulements souterrains complexes, pages 207--213.
\newblock Elsevier, 1997.

\bibitem{Caswell1992}
H.~Caswell and A.~John.
\newblock {\em Individual-based models and approaches in ecology}, chapter From
  the individual to the population in demographic models, pages 36--66.
\newblock Chapman et Hall, 1992.

\bibitem{Crutchfield1992}
J.~Crutchfield.
\newblock {\em Nonlinear Dynamics of Ocean Waves}, chapter Discovering Coherent
  Structures in Nonlinear Spatial Systems, pages 190--216.
\newblock World Scientific., Singapore, 1992.

\bibitem{DeAngelis1992}
D.~L. DeAngelis and L.~J. Gross, editors.
\newblock {\em Individual-Based Models and Approaches in Ecology: Populations,
  Communities and Ecosystems}.
\newblock Chapman and Hall, 1992.

\bibitem{Drogoul1992}
A.~Drogoul and C.~Dubreuil.
\newblock {\em Decentralized Artificial Intelligence III}, chapter
  Eco-Problem-Solving : results of the N-Puzzle, pages 283--295.
\newblock North Holland, 1992.

\bibitem{Gasser1992}
L.~Gasser.
\newblock {\em Decentralized A.I}, chapter Boundaries, Identity and
  Agggregation : Pluralities issues in Multi-Agent Systems.
\newblock Elsevier, 1992.

\bibitem{Langton1986}
C.~Langton.
\newblock Studying artificial life with cellular automata.
\newblock {\em Physica D}, 22, 1986.

\bibitem{Leonard1980}
A.~Leonard.
\newblock Vortex methods for flow simulation.
\newblock {\em Journal of Computational Physics}, 37:289--335, 1980.

\bibitem{Lesieur1987}
M.~Lesieur.
\newblock {\em Turbulence in fluids}.
\newblock Martinus Nijhoff Publishers, 1987.

\bibitem{Marcenac1997}
P.~Marcenac.
\newblock Mod\'elisation de syst\`emes complexes par agents.
\newblock {\em Techniques et Sciences Informatiques}, 1997.

\bibitem{Palmer1992}
J.~Palmer.
\newblock {\em Individual-based models and approaches in ecology}, chapter
  Hierarchical and concurrent individual-based modelling, pages 36--66.
\newblock Chapman et Hall, 1992.

\bibitem{Servat1998}
D.~Servat, E.~Perrier, J.-P. Treuil, and A.~Drogoul.
\newblock When agents emerge from agents : Introducing multi-scale viewpoints
  in multi-agent simulations.
\newblock In {\em MABS}, pages 183--198, 1998.

\bibitem{Tranouez2003}
P.~Tranouez, S.~Lerebourg, C.~Bertelle, and D.~Olivier.
\newblock Changing the level of description in aquatic ecosystem models: an
  overview.
\newblock In {\em ESMc' 2003}, Naples, Italy, October 2003. ESMC.

\bibitem{Tranouez2005}
P.~Tranouez, G.~Pr\'eost, C.~Bertelle, and D.~Olivier.
\newblock Simulation of a compartmental multiscale model of predator-prey
  interactions.
\newblock {\em Dynamics of Continuous, Discrete and Impulsive Systems Journal},
  2005.

\bibitem{VonNeuman1966}
J.~VonNeuman and A.~Burks.
\newblock Theory of self-reproduction automata.
\newblock University of Illinois Press, 1966.

\end{thebibliography}

\printindex
\end{document}